\documentclass[preprint]{JHEP3}
\usepackage{epsf}
\preprint{SOGANG-HEP 312/04 \\
          gr-qc/0404098}
\title{Wormhole phase in the RST model}
\author{Won Tae Kim and Edwin J. Son \\
         Department of Physics and Basic Science Research Institute,\\
         Sogang University, C.P.O. Box 1142, Seoul 100-611, Korea \\
         E-mail: \email{wtkim@sogang.ac.kr},
                 \email{sopp@string.sogang.ac.kr}}
\abstract{
We show that the RST model describing the exactly soluble black hole
model can have a dynamical wormhole solution along with an appropriate
boundary condition. The necessary exotic matter which is usually
negative energy density is remarkably produced by the quantization of
the infalling matter fields.
Then the asymptotic geometry in the past is two-dimensional anti-de
Sitter(AdS$_2$), which implies the
exotic matter is negative. 
As time goes on, the wormhole eventually evolves into the black hole
and its Hawking radiation appears.
The throat of the static RST wormhole is lower-bounded
but in the presence of infalling matter it collapses to a black hole.
}

\begin{document}
\section{Introduction}
\label{sec:intro}
A two-dimensional quantum gravity coupled to conformal matters has
been extensively studied \cite{cghs}. It is exactly soluble at the
classical level, however, the semiclassical equations of motion have
not been solved in a closed form. Then, the model has been modified by
a number of authors to allow explicit construction for exact
quantum black hole solutions. Among them, Russo, Susskind, and
Thorlacius (RST) \cite{rst} have added a specific term, rather than
modifying the dilaton potential as was done in
Refs.~\cite{rt}. They have constructed an exactly solvable
semiclassical model, which has semiclassical solutions describing the
formation and the evaporation of black holes with an asymptotic-flat
boundary condition.

On the other hand, 
there has been great interest in space-time wormholes
\cite{wheeler,mt} describing travels to other universes,
interstellar travels, and time machines, {\it etc}. Classically, in
order to construct a Lorentzian wormhole, the exotic matter 
violating Weak Energy Condition (WEC), Null Energy Condition (NEC),
Strong Energy Condition (SEC), and/or Dominant Energy Condition (DEC),
{\it etc} is required \cite{mty,visser}. Especially, for exactly
soluble wormhole models, the ghost fields have been used in obtaining
stable wormhole solutions \cite{hkl,khk,hky,ks}. 
However, if one considers the quantum theory rather than the
above-mentioned classical one, then 
it might be more plausible to get the exotic source from the quantum
regime since our universe is eventually governed by
quantum mechanics. 

So, in this paper, we would like to consider the energy-momentum
tensors from the nonlocal term generated by the one-loop effective
action in the RST model as a candidate for the exotic source, instead
of the artificial classical negative energy density. In fact, it has
been shown that the arbitrarily small violations of energy conditions
yield a wormhole solution \cite{vkd}.
Then, we find the wormhole phase 
different from the conventional black hole phase in the RST model.  
Since the quantum-corrected energy-momentum tensors are sensitive to
the boundary condition, we can use appropriate energy-momentum tensors
satisfying the constraint equations for the wormhole geometry.
Thus, the quantum-mechanically induced negative energy density
produces the static wormhole, and as time goes 
on it eventually evolves into a black hole.

We recast the RST model to obtain the general solution, in section
\ref{sec:sol}, and present a wormhole solution by imposing the
static-wormhole boundary condition at the earliest time, which yields
an asymptotic-AdS geometry. In section \ref{sec:exoticity}, the
exoticity function \cite{mt} is evaluated for our model and the
asymptotic behavior of the quantum correction of the stress-energy is
presented. Finally, summary and discussions are given in section
\ref{sec:discussion}.

\section{Wormhole solution in the RST model}
\label{sec:sol}
It has been well-known that the RST model gives evaporating black
hole or Minkowskian vacuum solutions depending on the boundary
conditions, however, other geometries have not been considered
yet. So one might think the other kind intriguing geometries,
specifically, wormhole soulutions by choosing the other appropriate
boundary conditions. Of course, the essential motive is to obtain the
wormhole solution without the {\it ad hoc} classical negative energy
density. We will show that the RST model naturally involves our
desirable wormhole solution, which is a big difference from the
previous works \cite{hkl,khk,hky,ks}.

We begin with the two-dimensional dilaton gravity action coupled to
conformal fields with the following quantum correction term in two
dimensions,
\begin{eqnarray}
S &=& S_{DG} + S_f + S_{qt}, \label{action} \\
S_{DG} &=& \frac{1}{2\pi} \int d^2 x \sqrt{-g} e^{-2\phi} \left[ R +
  4(\nabla\phi)^2 + 4\lambda^2 \right], \label{action:gravity} \\
S_f &=& \frac{1}{2\pi} \int d^2 x \sqrt{-g} \sum_{i=1}^N \left[ -
  \frac{1}{2} \left( \nabla f_i \right)^2 \right],
  \label{action:matter} \\
S_{qt} &=& \frac{1}{2\pi} \int d^2 x \sqrt{-g} \left[ -
  \frac{\kappa}{2} \phi R - \frac{Q^2}{2} R \frac{1}{\Box} R \right],
\label{action:quantum}
\end{eqnarray}
where $g$, $\phi$, and $f_i$'s are the metric, the dilaton field, and
the conformal matter fields, and $\lambda^2$ is a cosmological
constant. Choosing $\kappa = 2 Q^2 = \frac{N - 24}{12}$ yields the well-known RST
model, where $N$ is the number of conformal matter fields,
which is taken to be large. This model gives a dynamical black hole
solution with the positive Arnowitt-Deser-Misner (ADM) mass \cite{rst}.

In the conformal gauge defined by $g_{+-} = - \frac{1}{2} e^{2\rho},
g_{--} = g_{++} = 0$, where $x^\pm = (x^0 \pm x^1)$, the above action
is written as \cite{rst}
\begin{equation}
S = \frac{1}{\pi} \int d^2 x \left[ - \partial_+ \chi \partial_- \chi
  + \partial_+ \Omega \partial_- \Omega + \lambda^2
  e^{\frac{2}{\sqrt{\kappa}} \left( \chi - \Omega \right)} + \frac12
  \sum_i \partial_+ f_i \partial_- f_i \right] \label{conf:action}
\end{equation}
with the constraints
\begin{equation}
\kappa t_\pm (x^\pm) = \sqrt{\kappa} \partial_\pm^2 \chi -
  \partial_\pm \chi \partial_\pm \chi + \partial_\pm \Omega \partial_\pm
  \Omega + \frac12 \sum_i \partial_\pm f_i \partial_\pm f_i,
  \label{conf:cons}
\end{equation}
where the functions $t_\pm(x^\pm)$ reflect the nonlocality of the
conformal anomaly of the action (\ref{action:quantum}), and we
performed a field redefinition to a Liouville theory \cite{liberati}:
$\chi = \sqrt{\kappa} \rho - \sqrt{\kappa} \phi / 2 + e^{-2\phi} /
\sqrt{\kappa}$ and $\Omega = \sqrt{\kappa} \phi / 2 + e^{-2\phi} /
\sqrt{\kappa}$, where we assumed $\kappa > 0$.
Then, the equations of motion are obtained from the action
(\ref{conf:action}),
\begin{eqnarray}
& & 2 \partial_+ \partial_- \chi + \frac{2}{\sqrt{\kappa}} \lambda^2
  e^{\frac{2}{\sqrt{\kappa}} \left( \chi - \Omega \right)} = 0,
  \label{conf:chi} \\
& & - 2 \partial_+ \partial_- \Omega - \frac{2}{\sqrt{\kappa}}
  \lambda^2 e^{\frac{2}{\sqrt{\kappa}} \left( \chi - \Omega \right)} =
  0, \label{conf:Omega} \\
& & \partial_+ \partial_- f_i = 0, \label{conf:matter}
\end{eqnarray}
and the key ingredient of the exact solubility is due to
Eqs.~(\ref{conf:chi}) and (\ref{conf:Omega}). Combining these, we
obtain $\partial_+ \partial_- \left( \chi - \Omega \right) = 0$, and
then the residual symmetry can be fixed by choosing $\chi = \Omega$
(or $\rho = \phi$) in the Kruskal gauge.
The general solutions of dilaton and conformal matter are obtained as
\begin{eqnarray}
\Omega (x^+, x^-) &=& - \frac{\lambda^2}{\sqrt{\kappa}} x^+ x^- +
  a_+(x^+) + a_-(x^-), \label{gen:dilaton} \\
f_i(x^+,x^-) &=& f_+^{(i)}(x^+) + f_-^{(i)}(x^-), \label{gen:matter}
\end{eqnarray}
where $a_\pm(x^\pm)$ are integration functions determined by the
constraints (\ref{conf:cons}) as
\begin{equation}
\kappa t_\pm = \sqrt{\kappa} \partial_\pm^2 a_\pm + \frac12 \sum_i
  \partial_\pm f^{(i)}_\pm \partial_\pm
  f^{(i)}_\pm. \label{kruskal:cons} 
\end{equation}
Integrating the constraints (\ref{kruskal:cons}), the solution is
obtained as
\begin{eqnarray}
\Omega (x^+, x^-) &=& - \frac{\lambda^2}{\sqrt{\kappa}} x^+ x^- + \int
  dx^+ \int dx^+ \left( \sqrt{\kappa} t_+ - \frac{1}{2\sqrt{\kappa}}
  \sum_i \partial_+ f^{(i)}_+ \partial_+ f^{(i)}_+ \right) \nonumber
  \\
& & + \! \int \! \! dx^- \! \! \int \! \! dx^- \! \left( \! \!
  \sqrt{\kappa} t_- \! - \!
  \frac{1}{2\sqrt{\kappa}} \sum_i \partial_- f^{(i)}_- \partial_-
  f^{(i)}_- \! \right) \! \! + \! C_+ x^+ \! + \! C_- x^- \! + \! D,\label{omegasol}
\end{eqnarray}
where $C_\pm$, and $D$ are integration constants.
Formally, the apparent horizon curves are given as
\begin{equation}
0 = \partial_\pm \Omega
  = C_\pm - \frac{\lambda^2}{\sqrt{\kappa}} x_{\mathrm h}^\mp +
  \int^{x_{\mathrm h}^\pm} dx^\pm \left( \sqrt{\kappa} t_\pm -
  \frac{1}{2\sqrt{\kappa}} \sum_i \partial_\pm f^{(i)}_\pm
  \partial_\pm f^{(i)}_\pm \right).
\end{equation}
Note that the evaporating black hole solution appears \cite{rst} 
within the assumption that the asymptotic geometry is described by the
Minkowskian, which eventually fixes the boundary functions as
 $t_\pm = 1 / 4 (x^\pm)^2$. 

With the exact solubility maintaining,
one might think of the other kind of the geometry, especially,
wormhole phase which will be eventually incorporated in this model. 
There exist largely two kinds of intriguing objects such as black hole and
wormhole depending on the boundary 
conditions. The latter case requires that the exotic matter source
is additionally present near the throat of the wormhole, and this exotic
source of the negative energy density can be produced by choosing
some different boundary conditions.   
Explicitly, the wormhole boundary condition means that the past and the future
horizon curves are coincident with each other, which is consistent
with the flaring-out condition in Ref.~\cite{mt}. In particular, the
two horizons are coincident at $x^+ = x^-$ for the static
wormhole. 

\begin{figure}
\begin{center}
\leavevmode
\epsfxsize=0.7\textwidth
\epsfbox{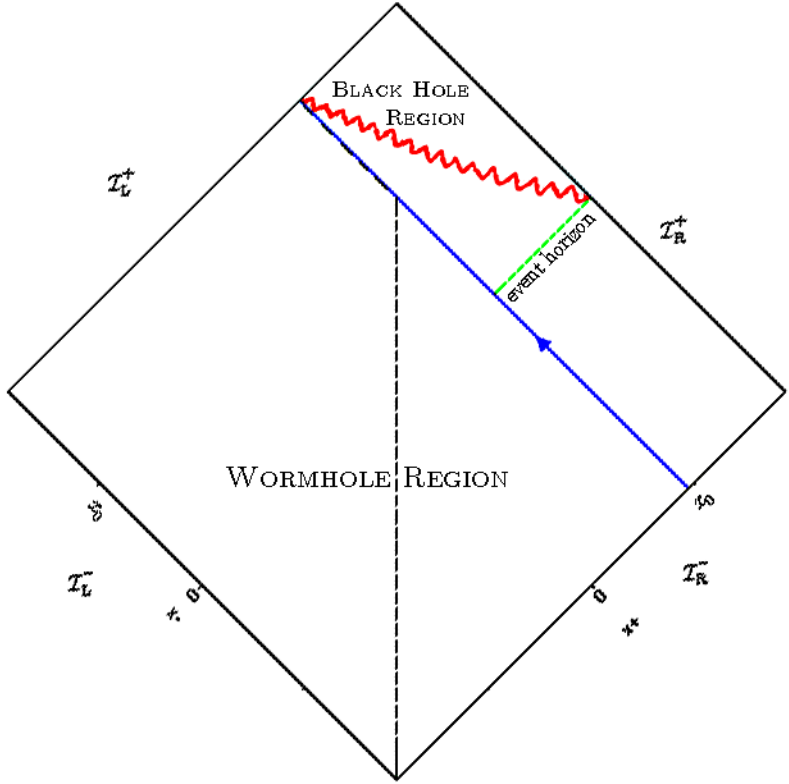}
\end{center}
\caption{The Penrose diagram of the wormhole with the infalling matter source
  simply for $M = x_0 = \lambda = \kappa = A = 1$ shows that the
  conformal matter splits degenerate horizons. The
  thick wiggly line representing the curvature singularity is
  cloaked by the event horizon.}
\label{fig:diagram}
\end{figure}

Returning to our specific model, we now consider 
the infalling matter source as $T_{++}^f=\frac12 \sum \partial_+
f^{(i)}_+ \partial_+ f^{(i)}_+ = A \delta ( x^+ - x_0 )$ and $T_{--}^f=\frac12
\sum \partial_- f^{(i)}_- \partial_- f^{(i)}_- = 0$, where $A$ is a
positive constant. 
If we require an appropriate boundary condition making the
wormhole in the past, then
the past and future horizon curves are coincident with the line $x^+ = x^-$ at
$x^\pm < x_0$ such that the relation $x_{\mathrm h}^+ = x_{\mathrm h}^-$ can
be used. Furthermore, requiring no radiation condition along ${\mathcal I}_R^-$
(${\mathcal I}_L^-$), similarly to the RST black holes at
$x^+ > x_0$ ($x^- > x_0$) \cite{rst}, we can fix $C_\pm$ and $t_\pm$ to  
\begin{equation}
C_\pm =
\lambda^2 x_0 / \sqrt{\kappa},~~t_\pm(x^\pm) = \frac{\lambda^2}{\kappa} \theta(x_0 - x^\pm),
\end{equation}
where $\theta(x)$ is 1 for $x>0$ and 0 for $x<0$. Then, the solution (\ref{omegasol})
is expressed by
\begin{eqnarray}
\Omega &=& D + \frac{\lambda^2}{\sqrt{\kappa}} x_0^2 -
  \frac{\lambda^2}{\sqrt{\kappa}} \left( x^+ - x_0 \right) \left( x^-
  - x_0 \right) - \frac{A}{\sqrt{\kappa}} \left( x^+ - x_0 \right)
  \theta \left( x^+ - x_0 \right) \nonumber \\
& &  + \frac{\lambda^2}{2\sqrt{\kappa}} \left( x_0 - x^+ \right)^2
  \theta \left( x_0 - x^+ \right) + \frac{\lambda^2}{2\sqrt{\kappa}}
  \left( x_0 - x^- \right)^2 \theta \left( x_0 - x^- \right),
  \label{sol:dilaton}
\end{eqnarray}
and the resulting horizon curves are from Eq. (\ref{sol:dilaton}),
\begin{eqnarray}
0 &=& \partial_+ \Omega (x^+_{\mathrm h}, x^-_{\mathrm h}) \nonumber \\
&=& - \frac{\lambda^2}{\sqrt{\kappa}} \left( x^-_{\mathrm h}
  - x_0  \right) - \frac{A}{\sqrt{\kappa}} \theta \left( x^+_{\mathrm h} - x_0
  \right) - \frac{\lambda^2}{\sqrt{\kappa}} \left( x_0 - x^+_{\mathrm h} \right)
  \theta \left( x_0 - x^+_{\mathrm h} \right), \\
0 &=& \partial_- \Omega (x^+_{\mathrm h}, x^-_{\mathrm h}) \nonumber \\
&=& - \frac{\lambda^2}{\sqrt{\kappa}} \left( x^+_{\mathrm h} - x_0  \right) -
  \frac{\lambda^2}{\sqrt{\kappa}} \left( x_0 - x^-_{\mathrm h} \right) \theta
  \left( x_0 - x^-_{\mathrm h} \right).
\end{eqnarray}
Note that the two horizons are splitted by the infalling matter,
which means that future horizon among the
coincident horizons for the wormhole throat is getting larger 
due to the infalling positive energy. Before the infalling matter,
the two horizons are still coincident so that the two universes are 
connected through the wormhole throat. However, after infalling,
the wormhole becomes unstable and it evolves into the black hole
described by the future event horizon, and its curvature singularity
is cloaked by the horizon. Therefore, the wormhole
disappears, which is shown in Fig.~\ref{fig:diagram}.

Next, to find out the meaning of $D$,  
let us define the curvature scalar $R$ as \cite{rst},
\begin{equation}
R = \frac{8e^{-2\phi}}{\Omega'} \left[ \partial_+ \partial_- \Omega -
  \frac{\Omega''}{\Omega'^2} \partial_+ \Omega \partial_- \Omega
  \right] \label{curvature}
\end{equation}
in the Kruskal gauge, where $' = d/d\phi$. 
Then, the singularity curve
is obtained from $\Omega ' = 0$ as $\Omega \left( x_{\mathrm s}^+,
  x_{\mathrm s}^- \right) = \sqrt{\kappa} \left( 1 - \ln (\kappa/4)
\right) / 4$ which becomes $0 = D + \lambda^2 x_0^2 / \sqrt{\kappa} -
\sqrt{\kappa} \left( 1 - \ln (\kappa/4) \right) / 4 + \lambda^2 \left(
  x_{\mathrm s}^+ - x_{\mathrm s}^- \right)^2 / 2 \sqrt{\kappa}$ in a
stable wormhole geometry ($x^\pm < x_0^\pm$), where the subscript
denotes the singularity. 
Because there should not exist any singularities
in a stable wormhole geometry, the constant $D$ is restricted by the
following condition,
\begin{equation}
D + \frac{\lambda^2}{\sqrt{\kappa}} x_0^2 - \frac{\sqrt{\kappa}}{4}
  \left( 1 - \ln \frac{\kappa}{4} \right) = \frac{M}{\lambda} >
  0,
\end{equation}
where $M$ is a parameter related to the wormhole throat.
Especially, for $M=0$, the singularity
appears along the line $x_{\mathrm s}^+ = x_{\mathrm s}^-$. 
As a result, eliminating the above $D$, the solution (\ref{sol:dilaton}) becomes 
\begin{eqnarray}
\Omega &=& \frac{M}{\lambda} + \frac{\sqrt{\kappa}}{4} \left( 1 - \ln
  \frac{\kappa}{4} \right) - \frac{\lambda^2}{\sqrt{\kappa}} \left(
  x^+ - x_0 \right) \left( x^- - x_0 \right) - \frac{A}{\sqrt{\kappa}}
  \left( x^+ - x_0 \right) \theta \left( x^+ - x_0 \right) \nonumber
  \\
& &  + \frac{\lambda^2}{2\sqrt{\kappa}} \left( x_0 - x^+ \right)^2
  \theta \left( x_0 - x^+ \right) + \frac{\lambda^2}{2\sqrt{\kappa}}
  \left( x_0 - x^- \right)^2 \theta \left( x_0 - x^- \right).
\end{eqnarray}
In the region of $x^\pm < x_0$, a static wormhole solution is given as
\begin{equation}
\Omega = \frac{M}{\lambda} + \frac{\sqrt{\kappa}}{4} \left( 1 - \ln
  \frac{\kappa}{4} \right) + \frac{\lambda^2}{2\sqrt{\kappa}} \left(
  x^+ - x^- \right)^2.
  \label{sol}
\end{equation}

By using this solution, we profile the curvature scalar and evaluate
the size of wormhole throat.
First of all, we define the dilaton field in terms of this
wormhole solution through the inverse function,  
\begin{equation}
\phi = \frac{1}{2\sqrt{\kappa}} \left[ 4 \Omega + \sqrt{\kappa} W (
  \xi ) \right], \label{rel:phi}
\end{equation}
where $W(\xi)$ satisfies $W(\xi) e^{W(\xi)} = \xi$, $\xi = -(4/\kappa)
e^{-4 \Omega / \sqrt{\kappa}}$, and $W(\xi) < 0$. 
Then, $\Omega'$ and
$\Omega''$ are obtained as $\Omega' = \frac12 \sqrt{\kappa} \left[ 1 +
  W( \xi ) \right]$ and $\Omega'' = - \sqrt{\kappa} W( \xi )$, and
$e^{-2\phi} = \sqrt{\kappa} \Omega'' / 4$.
Thus, the curvature scalar
(\ref{curvature}) becomes 
\begin{equation}
R = \frac{4 W(\xi)}{1+W(\xi)} \left[ \lambda^2 + \frac{4 \lambda^4
  W(\xi)}{\kappa (1+W(\xi))^2} \left( x^+ - x^- \right)^2 \right].
\end{equation}
Since $-1/e \le \xi < 0$, we naturally have two types of solutions for
$W(\xi)$; one corresponds to the weak (string) coupling, $W(\xi) \le
-1$, and the other corresponds to the strong coupling, $-1 \le W(\xi)
< 0$, where the coupling is defined as $g = e^\phi$, and we confine
the weak coupling case \cite{cghs}.
In the asymptotic region, $| x^+ - x^- | \to \infty$, the solution
$\Omega$ diverges and $\xi$ exponentially converges to zero, then
$W(\xi) \to - \infty$. Using the following approximation, $W(\xi)
\approx \ln (-\xi) \approx - 2\lambda^2 (x^+ - x^-)^2 / \kappa$, the
curvature scalar becomes $R_{\mathrm{asym}} \rightarrow
-4\lambda^2$. Thus, asymptotically AdS$_2$ spacetime is obtained. 

On the other hand, 
near the throat, $| x^+ - x^- | \to 0$, $\xi$ converges as $\xi_0 =
-(1/e) e^{-4 M / \lambda \sqrt{\kappa}}$, then $W(\xi)$ becomes
$W_0 = W(\xi_0)$. Along with the approximation of the solution $\Omega$ 
of $\xi = \xi_0 \exp \left[- \frac{2\lambda^2}{\kappa} \left( x^+ - x^-
\right)^2 \right] \approx \xi_0 \left[ 1 - \frac{2\lambda^2}{\kappa}
\left( x^+ - x^- \right)^2 \right]$ in this region,  
the function $\xi$ can be approximately written as
\begin{eqnarray}
\xi &=& W(\xi) e^{W(\xi)} \nonumber \\
&\approx& \xi_0 + \xi_0 \frac{1+W_0}{W_0} \left[ W(\xi) - W_0 \right]
  + \xi_0 \frac{2+W_0}{2W_0} \left[ W(\xi) - W_0 \right]^2,
  \label{approx}
\end{eqnarray}
where we expanded around $W_0$ and used the relation $e^{W_0} = \xi_0
/ W_0$. For $M>0$, $W(\xi)
\approx W_0 - W_0 ( 1 - \xi / \xi_0 ) / ( 1 + W_0 )$ in the leading
order of Eq.~(\ref{approx}), and then $R$ is finite,
$R_{\mathrm{throat}} \rightarrow 4 \lambda^2 W_0 / (1 + W_0)$.

As for the size of the wormhole throat, 
$e^{-2\phi}$ is assumed to be analogously related to
higher-dimensional radial coordinate, and 
can be used to check whether the wormhole is closed or
not, which has already been used in two-dimensional physical quantity \cite{strominger}. 
From Eq. (\ref{rel:phi}),
it is written as
\begin{equation}
e^{-2\phi} = - \frac{\kappa}{4} W(\xi) \equiv \lambda^2 r^2,
\end{equation}
where $r$ is a radial coordinate satisfying
\begin{equation}
r^2 = - \frac{\kappa}{4\lambda^2} W(\xi) \ge -
  \frac{\kappa}{4\lambda^2} W_0 \ge \frac{\kappa}{4\lambda^2}. \label{radial}
\end{equation}
Then, the minimal throat radius is given as $r^2 = - \kappa W_0 / 4
\lambda^2$. 
Note that it is quantum-mechanical because $\kappa$ involves Plank constant. 

\section{Exoticity of quantum source}
\label{sec:exoticity}
Now, it seems to be appropriate to mention how
the result describing the wormhole geometry (\ref{sol}) comes out
by directly calculating the well-known exotic condition 
expressed by exoticity function $\zeta$ \cite{mt}. To do
so, we change the coordinate system to ``the proper reference frame''
(the hatted coordinate system),
\begin{eqnarray*}
ds^2 &=& - e^{2\rho(x)} dx^+ dx^- \\
&=& - \exp \left( 2 \rho \pm \lambda \sigma^+ \pm \lambda \sigma^- \right)
d\sigma^+ d\sigma^- \\
&=& - \left( d\sigma^{\hat{0}} \right)^2 + \left( d\sigma^{\hat{1}}
\right)^2.
\end{eqnarray*}
where $\sigma^\pm = \sigma^0 \pm \sigma^1$.
Then, in this coordinate system, 
the ($\hat{0}$,$\hat{0}$)- and
($\hat{1}$,$\hat{1}$)-components of the energy-momentum tensors
$T_{\mu\nu} = <T_{\mu\nu}^f>$ are obtained 
from Eq. (\ref{action:quantum}) as
\begin{eqnarray}
T_{\hat{0} \hat{0}} &=& 2 \lambda^2 C_0 \left( \frac{3 - W_0}{1 + W_0}
\right), \\
T_{\hat{1} \hat{1}} &=& -2 \lambda^2 C_0,
\end{eqnarray}
in the leading order of $\sigma^1$, near the throat, $|x^+ - x^-| \to
0$. Note that $C_0 = \exp[- 4M/\lambda \sqrt{\kappa} - 1 +
\ln(\kappa/4) - W_0]$, and Eqs.~(\ref{rel:phi}) and (\ref{approx})
were used. So, the exoticity function reads as
\begin{equation}
\zeta \equiv \frac{-T_{\hat{1} \hat{1}} - T_{\hat{0}
    \hat{0}}}{|T_{\hat{0} \hat{0}}|} = \frac{4 \lambda^2
    C_0}{|T_{\hat{0} \hat{0}}|} \left( \frac{W_0 - 1}{W_0 + 1} \right)
    > 0,
\end{equation}
since $W_0 < -1$.
Thus, the quantum-mechanically induced energy plays the role of the
exotic source since the exoticity is satisfied at the throat.
In fact, {\it ab initio},
the classical exotic source is usually considered for 
lower-dimensional wormhole models \cite{hkl,khk,hky,ks}. 
In  Ref.~\cite{hkl}, the classical wormhole is given by the 
classical dilaton gravity action coupled to ghost fields with
wrong sign kinetic term to construct wormholes. With the
great help of the ghost field, 
the desirable classical wormhole solution is
obtained, however, it seems to be awkward in that the
origin of the exotic source is more or less arbitrary.  
In the present model, we exploited the quantum effect 
in order to produce the negative energy density which results in
the wormhole. 

Finally, we comment on the quantum radiation from the black hole,
and it is explicitly calculated as
$< T^f_{--} > =2 \sqrt{\kappa} \partial_-^2 \Omega /(1 +
W(\xi)) - 4 (\partial_- \Omega)^2 [(1 - W(\xi)) / (1 + W(\xi))^3] -
\kappa t_-$, where Eq.~(\ref{rel:phi}) is used.
At the spatial infinity,
${\mathcal I}^+_R$, it is simply reduced to
$< T^f_{--} >_{{\mathcal I}^+_R} \to - \lambda^2 \theta(x_0 - x^-) +
\kappa / 4 ( x^- - x_0 + A/\lambda^2)^2$. Because of the first term,
this might be understood by
the (anti-)evaporation \cite{bh}, however, it is caused by the
energy-momentum tensor, $< T^f_{--} >_{{\mathcal
I}^-_L} \to - \lambda^2 \theta(x_0 - x^-)$
at the infinity ${\mathcal I}^-_L$.
Thus, the background energy-momentum contributing to construct a
wormhole is not a real radiation so that the net radiation is
$< T^f_{--} >_{{\mathcal I}^+_R} - < T^f_{--} >_{{\mathcal I}^-_L} =
\kappa / 4 ( x^- - x_0 + A/\lambda^2)^2$.
Note that at the past null infinities the geometry was AdS$_2$,
which is related to the fact that the energy momentum tensor is
constant, $-\lambda^2$ at ${\mathcal I}^-_L$.

\section{Discussion}
\label{sec:discussion}
We have shown that the RST model has a wormhole solution along with
some boundary conditions so that the wormhole can be constructed
without introducing an exotic matter by hand.
It is interesting to note that the throat of a static quantum
wormhole (\ref{radial}) is lower-bounded when the wormhole is singularity-free. 
This means that it is always open
while the throat of the classical wormhole can be closed \cite{hkl}. 
In our model, the quantum wormhole collapses to the black hole, and then the
infalling classical matter
have been trapped in the black hole. So, it will be interesting
to study how to overcome this singularity for traversability of the
wormhole for further work.

\acknowledgments
We would like to thank John J. Oh for exciting discussions.
This work was supported by Grant No. 2000-2-11100-002-5 from the Basic
Research Program of the Korean Science and Engineering Foundation.

\end{document}